\documentclass[11pt, a4paper]{article}

\usepackage{mathtools}

\usepackage{amsmath}
\usepackage{amsfonts}
\usepackage{amssymb}
\usepackage{bbold}
\usepackage{mathrsfs}
\usepackage{latexsym} 
\usepackage{cancel}
\usepackage{bm}
\usepackage{graphicx, rotating}
\usepackage{epstopdf}
\usepackage{epsfig}
\usepackage{latexsym}
\usepackage{color}
\usepackage[dvipsnames]{xcolor}
\usepackage{cite}
\usepackage{slashed}
\usepackage{hyperref}
\usepackage{comment}
\usepackage[utf8]{inputenc}
\usepackage{soul}
\usepackage{multirow}
\usepackage{graphics,subfigure}
\usepackage[dvipsnames]{xcolor}
\usepackage{physics}

\hypersetup{colorlinks, citecolor=bluscuro, linkcolor=black, urlcolor=bluscuro}
\definecolor{bluscuro}{rgb}{0.15, 0.2, .85}

\newcommand{\be}{\begin{equation}}
\newcommand{\ee}{\end{equation}}
\newcommand{\bea}{\begin{eqnarray}}
\newcommand{\eea}{\end{eqnarray}}

\newcommand\blfootnote[1]{%
  \begingroup
  \renewcommand\thefootnote{}\footnote{#1}%
  \addtocounter{footnote}{-1}%
  \endgroup
}

 \def\bea{\begin{eqnarray}}
  \def\eea{\end{eqnarray}}

	\def \beq {\begin{equation}}
	\def \eeq {\end{equation}}
	\def \ba {\begin{array}}
	\def \ea {\end{array}}

\newcommand{\orcid}{\includegraphics{Fig/orcid.pdf}}
\newcommand{\orcidlink}[1]{\href{https://orcid.org/#1}{{\orcid}}}

\begin{document}

\begin{titlepage}
\begin{flushright}
IFT-UAM/CSIC-22-113
\end{flushright}
\begin{center} ~~\\
\vspace{0.5cm} 
\Large {\bf\Large Testing  entanglement and Bell inequalities  in $H \to ZZ$} 
\vspace*{1.5cm}

\normalsize{
{\bf 
J. A. Aguilar-Saavedra
\blfootnote{ja.a.s@csic.es}, A. Bernal
\blfootnote{
alexander.bernal@csic.es},
J. A. Casas
\blfootnote{j.alberto.casas@gmail.com
} and
J. M. Moreno
\blfootnote{
jesus.moreno@csic.es}
 } \\
 
\smallskip  \medskip
{\it Instituto de F\'\i sica Te\'orica, IFT-UAM/CSIC,}\\
\it{Universidad Aut\'onoma de Madrid, Cantoblanco, 28049 Madrid, Spain}}

\medskip

\vskip0.6in 

\end{center}

\centerline{ \large\bf Abstract }
\vspace{.5cm}

We discuss quantum entanglement and violation of Bell inequalities in the $H\rightarrow ZZ$ decay, in particular when the two $Z-$bosons decay into light leptons. Although such process implies an important suppression of the statistics, this is traded by clean signals from a `quasi maximally-entangled' system, which makes it very promising to check these crucial phenomena at high energy. In this paper we devise a novel framework to extract from $H \to ZZ$ data all significant information related to this goal, in particular spin correlation observables. In this context we derive sufficient and necessary conditions for entanglement in terms of only two parameters. Likewise, we obtain a sufficient and improved condition for the violation of Bell-type inequalities. The numerical analysis shows that with a luminosity of $L = 300~\text{fb}^{-1}$ entanglement can be probed at $> 3\sigma$ level. For $L = 3~\text{ab}^{-1}$ (HL-LHC) entanglement can be probed beyond the $5\sigma$ level, while the sensitivity to a violation of the Bell inequalities is at the $4.5\sigma$ level.

\vspace*{2mm}
\end{titlepage}


\section{Introduction}
\label{sec:Intro}

Entanglement is possibly the aspect of quantum mechanics that shows the greatest departure from classical conceptions 
\cite{Schrodinger:1935}. Such departure is evidenced by the 
violation of Bell inequalities \cite{Bell:1964kc} by quantum mechanics, something unfeasible in any theory consistent with the classical notions of locality and realism. Let us recall here that entanglement is a necessary but not sufficient condition for the violation of the Bell inequalities.
In consequence, it is highly relevant to test both phenomena at different scales, in particular at the highest possible energies
\cite{Tornqvist:1980af,Abel:1992kz}.

This objective has recently been explored in several articles \cite{Afik:2020onf, Fabbrichesi:2021npl, Severi:2021cnj,Aoude:2022imd, Afik:2022kwm,Aguilar-Saavedra:2022uye}, on the top-antitop system ($t\bar t$) at the LHC. On the other hand, a natural arena for these tests is provided by the Higgs boson decays in various channels. Certainly, the statistics is much smaller than for $t\bar t$ production, but the physical system is much closer to a maximally entangled state.  A first investigation in this sense was carried out in ref.\cite{Barr:2021zcp}, considering the decay of the Higgs boson ($H$) into $W^+W^-$.

In this paper we will mainly focus on the $H\rightarrow ZZ$ decay, in particular when the two $Z-$bosons decay into light leptons. Admittedly, this amounts to an important suppression of the statistics, which is traded by clean signals from a `quasi maximally-entangled' system. On the other hand, an important aspect in this kind of challenge is to devise a framework to easily extract from $H \to ZZ$ data all significant information related
to entanglement and the violation of Bell inequalities, in particular the 80 spin and spin correlation observables.
Then we study necessary and sufficient conditions for entanglement and violation of Bell inequalities in terms of observable quantities and analyze the feasibility of these checks by using actual experimental data. This represents the main goal of the paper.

In section \ref{sec:generalditions} we review the definition and conditions for quantum entanglement and Bell inequalities, focussing on a system with two dimension-3 subsystems. 
In section \ref{sec:HZZ} we formulate the spin density matrix $\rho$ associated to the $ZZ$ system that arises from Higgs decays. We describe there the constraints on $\rho$ from symmetry considerations and express the matrix in an appropriate basis for Hermitian operators, with coefficients that can be determined from experimental data. In section \ref{sec:EntConds} we give sufficient and necessary conditions for entanglement in the $\rho$ matrix, expressing them in terms of the above coefficients. In section \ref{sec:BellConds} we perform a similar task for the conditions for the violation of Bell inequalities, introducing also a new Bell operator which is a more powerful indicator of that violation than other proposals in the literature. In section \ref{sec:numerical}
we investigate  the statistical sensitivity of future experimental measurements to the above described entanglement and violation of Bell inequalities.
We will show that there are good prospects to probe entanglement at the $3\sigma$ level at
the LHC with the Run 2+3 combined luminosities. At the HL-LHC, that significance would be well above $5 \sigma$, and the violation of Bell inequalities would be probed at the $4.5\sigma$ level.
Finally, in section \ref{sec:conclusions} we summarize this work and present our conclusions.
The appendix is devoted to expound the general construction of the Bell operator for the CGLMP inequality.

\section{Conditions for entanglement and violation of Bell inequalities}\label{sec:generalditions}

By definition, a state of two subsystems (Alice and Bob) is entangled if it is not separable, i.e. if its density matrix cannot be expressed as
\bea
\rho_{\rm sep} = \sum_n p_{n}\, \rho_n^{A} \otimes \rho_n^B\ ,
\label{rhosep}
\eea
where ${p_n}>0$ are classical probabilities, with $\sum p_n = 1$, and $\rho_n^{A}, \rho_n^B$ are density matrices acting in the Alice and Bob Hilbert spaces. A general test for a density matrix to determine whether it corresponds to a separable or an entangled state is not known. The most popular one is the Peres-Horodecki criterion \cite{Peres:1996dw, Horodecki:1997vt}, which provides a {\em sufficient} condition for entanglement: denoting $\ket{i}$, $\ket{\mu}$ two orthonormal bases of the  ${\cal H}_A$, ${\cal H}_B$ Hilbert spaces, and $\rho_{i\mu,j\nu}$ the density matrix of the global system, then a new matrix is constructed by transposing only the indices of Bob (or Alice), 
\bea
\rho^{T_2}=\rho_{i\nu,j\mu}\ .
\label{rhoT2}
\eea 
If  $\rho^{T_2}$ has at least one negative eigenvalue, then the $\rho$ matrix describes an entangled state. This sufficient condition is also necessary in two general cases: $dim\, {\cal H}_A= dim\, {\cal H}_B=2$ (qubits) and $ dim\, {\cal H}_A=2$, $dim\, {\cal H}_B=3$ (and vice-versa), but not for $ dim\, {\cal H}_A=dim\, {\cal H}_B=3$ (qutrits) or larger. This is the case of the spin states of the massive vector bosons. However, as we will see soon, for the $ZZ$ system stemming from a Higgs decay, the Peres-Horodecki is a {\em necessary} condition as well.

\vspace{0.2cm}
\noindent
Concerning the Bell inequalities, for subsystems of dimension 3, as the case at hand, several Bell-like relations have been explored. Typically, when the system is not very far from a maximally entangled state the popular CHSH inequality \cite{Clauser:1969ny}, which is optimal for qubits, does not provide the maximal departure from local realism predictions \cite{CGLMP2002}. A much more powerful relation is given by the so-called CGLMP inequality \cite{CGLMP2002}. Namely, if $A_1, A_2$ and $B_1, B_2$ are observables in ${\cal H}_A$, ${\cal H}_B$ that take (or are assigned to take) three possible values, $\pm 1, 0$ then the following inequality should hold in any local-realistic theory
\bea
I_3 &=& P\left(A_1=B_1\right)+P\left(B_1=A_2+1\right)+P\left(A_2=B_2\right)+P\left(B_2=A_1\right)\nonumber
\\
&&
-\left[P\left(A_1=B_1-1\right)+P\left(B_1=A_2\right)+P\left(A_2=B_2-1\right)+P\left(B_2=A_1-1\right)\right]\leq 2 \, ,
\label{CGLMP}
\eea
where $P\left(B_i=A_j+a\right)$ denotes the probability that the measurement of the observable $B_i$ gives the same result as the one of $A_j$ plus $a$ (mod. 3). 

In quantum mechanics the above probabilities are expressed in terms of expectation values of the appropriate projectors; e.g. $P\left(A_1=0\right)=\langle P^0_{A_1}\rangle=
\Tr{\rho\; P^0_{A_1} \otimes \mathbb{1}_3} $, where $P_{A_1}^0$ is the projector associated to the eigenvalue $0$ of $A_1$.
Consequently, the value of $I_3$ can be expressed as the expectation value of a certain operator, $I_3=\langle{\cal O}_{\text Bell}\rangle$, and the CGLMP inequality reads
\bea
I_3 = \langle {\cal O}_{\text Bell}\rangle\ = \Tr{\rho\; {\cal O}_{\text Bell}} \leq 2\ .
\label{I3}
\eea
The general expression of the ``Bell operator" ${\cal O}_{\text Bell}$, in terms of the four chosen observables, $A_1, A_2, B_1, B_2$, is given in detail in Appendix \ref{sec:AppendixA}.
Of course, in order to optimize the violation of the CGLMP-inequality (\ref{I3}) in quantum mechanics, a smart choice of the $A_1, A_2$, $B_1, B_2$ has to be made depending on the state $\rho$ at hand. This issue will be examined in detail in section \ref{sec:BellConds} below.

\section{The $H\rightarrow ZZ$ system}
\label{sec:HZZ}

\subsection{Expected form of the density matrix}
\label{sec:3.1}

The general spin state of the $ZZ$ system is described by a density operator, $\rho$, acting on the (dim 9) Hilbert space defined by
the three spin states of each $Z$. It is important to note that the propagator of the off-shell $Z$ boson also has a scalar component, whose contribution cancels when coupled to massless final state fermions, as is our case; therefore, we can safely consider the off-shell $Z$ boson as a spin-1 particle too~\cite{Groote:2012jq}.

We will work in the conventional basis of eigenstates of the  third component of the spin for each boson, $|+\rangle, |0\rangle, |-\rangle$, and in the center of mass (CM) reference system, defining the $z-$axis along the 3-momentum, $\vec k$, of one of the $Z$'s. Furthermore, to
avoid ambiguities, we will choose the latter as the $Z-$boson with the largest invariant mass, which is always well defined in the $H\rightarrow ZZ$ process. Note that with this choice of reference system, the sign of third component of spin coincides with the helicity
for one of the bosons (and minus the helicity for the other). 

It is interesting to discuss how far one can go to in determining the texture of the $\rho$ matrix, just based on the symmetries of the system. First of all, since in this case the two $Z$ bosons arise from the Higgs decay, the spin component along the momentum direction, i.e. $J_z$, is conserved and vanishing for the joint system. 
This means that only the 9 entries of the $\rho$ operator corresponding to $\langle u_i|\rho|u_j\rangle$, with $\ket{u_1}=\ket{+ -}, \ket{u_2}=\ket{00},\ket{u_3}=\ket{- +}$, can be different from zero.
In general, the $\rho$ operator will be a convex combination \bea
\rho=\sum p_\ell |\ell\rangle\langle \ell|,\ \ \ \ \ \ {\rm with}\ \ \    p_\ell\geq 0,\ \ \sum p_\ell=1 
\label{rho-l}
\eea
and
\bea
|\ell \rangle = \alpha_1 \ket{+-}+ \alpha_2 \ket{00} + \alpha_3\ket{-+} ,\ \ \ \ \ \ {\rm with}\  \sum_{i=1,2,3}|\alpha_i|^2=1 \, . 
\label{l-state}
\eea
Since in the $H\rightarrow ZZ$ process  parity is conserved, for a particular event the $ZZ$ state must be necessarily of the form
\bea
\ket{\psi_{ZZ}} = \frac{1}{\sqrt{2+\beta^2}}\left(\ket{+-} - \beta \ket{00} + \ket{-+}\right) \, ,
\label{psiZZ}
\eea 
with $\beta$ real.
Note that the $\alpha_1 = \alpha_3 = 0$ case in (\ref{l-state}) is recovered for $|\beta| \to \infty$.
In the rest frame the decay of the Higgs boson obviously has spherical symmetry, so the
value of $\beta$ just depends on the non-trivial kinematical variables, namely the two invariant masses of the vector bosons, $m_1, m_2$, and the modulus of the corresponding 3-momentum $|\vec k|$, which is related to the former by $\sqrt{|\vec k|^2+m_1^2} + \sqrt{|\vec k|^2+m_2^2}=m_H$.
We can get an accurate expression for $\beta$ using the Lorentz structure of the interaction term $\propto \eta_{\mu\nu}HZ^\mu Z^\nu$ in the SM.
Then the scalar state of the $ZZ$ system in the CM frame is given by \cite{Caban:2008qa}
\bea
\ket{\psi_{ZZ}}=\eta_{\mu\nu}\  e^\mu_\sigma(m_{1},\vec k)\  e^\nu_\lambda(m_2, -\vec k)\  |{\vec k, \sigma}\rangle_A |{-\vec k, \lambda}\rangle_B \, ,
\label{Lorentzdec}
\eea
where $\sigma, \lambda$ represent spin states
and  
\bea
e^\mu_\sigma(m,\vec k)=
\left.
\left(
\begin{array}{ccc}
 0 & \frac{|\vec k|}{m} & 0 
 \\
 -\frac{1}{\sqrt{2}} & 0 & \frac{1}{\sqrt{2}}
 \\
 \frac{i}{\sqrt{2}} & 0 & \frac{i}{\sqrt{2}}
 \\
 0 & -\frac{\sqrt{|\vec k|^2+m^2}}{m} & 0 
\end{array}
\right)\right.\, .
\eea 
Comparing Eqs.~(\ref{psiZZ}), (\ref{Lorentzdec}) we finally get
\bea
\beta=
1+\frac{m_H^2-(m_1+m_2)^2}{2 m_1 m_2}.
\label{x1}
\eea
Typically, the largest $Z-$mass, say $m_1$, is close to on-shell $m_Z$.
From Eq.~(\ref{x1}) we see that $\beta\geq 1$, with $\beta=1$ corresponding to the decay into two at-rest $Z$'s. In the latter case the spin-state of the $ZZ$ system is the singlet one, which is maximally entangled, see Eq.~(\ref{psiZZ}). Likewise, as $m_2$ decreases, and so $|\vec k|$ approaches its maximal value, the state goes to $\ket{00}$.
In consequence, the larger the mass of the off-shell $Z$, 
the larger the entanglement and the opportunities to experimentally show both entanglement and violation of Bell inequalities.

\begin{figure}[h!]
   \begin{center}
        \includegraphics[scale=0.45]{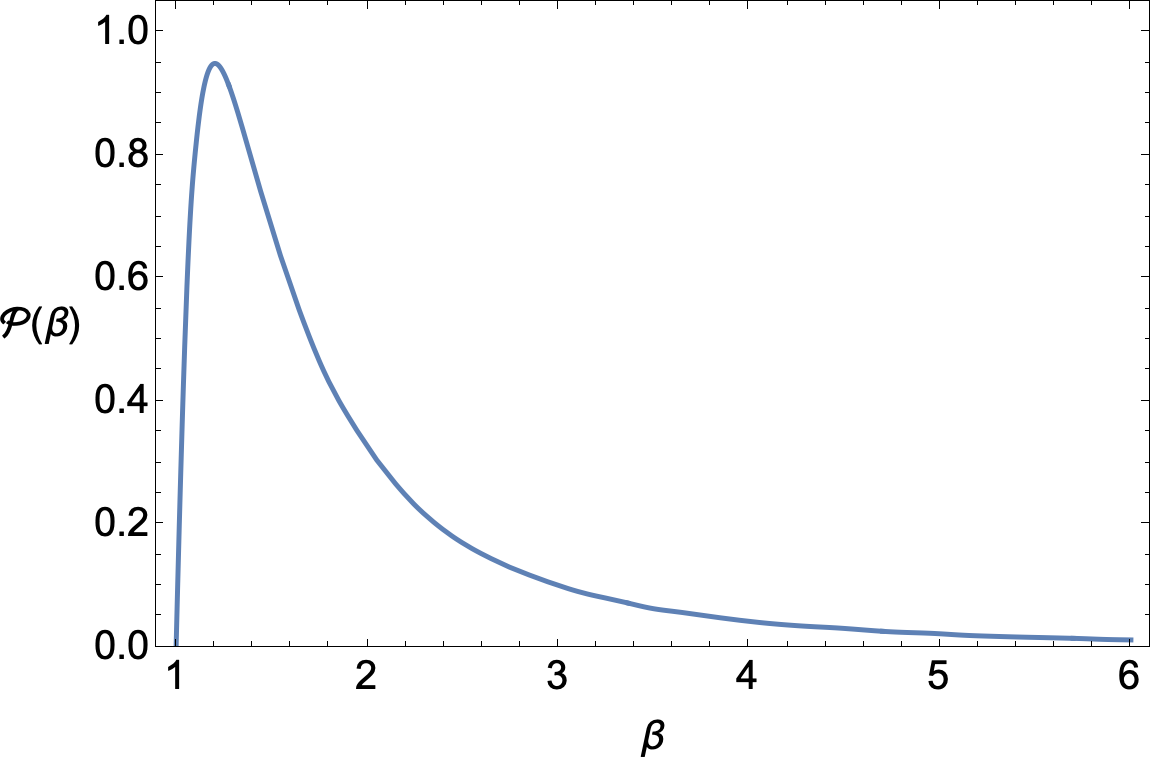}
     \end{center}
	\caption{ Probability distribution of $\beta$, see Eq.~(\ref{x1}), obtained with a Monte Carlo simulation when no cuts are implemented.}
    \label{fig:px}
\end{figure}

For a given value of $\beta$
the final spin state is pure and the $\rho$ matrix, say $\rho_{{}_\beta}$, is completely determined. However, when one gathers data from different kinematical configurations, the state becomes a
mixture
\bea
\rho=\int d\beta\ {\cal P}(\beta) \rho_{{}_\beta}\, .
\label{average}
\eea
Once the probability ${\cal P}(\beta)$ is known, the final $\rho$ becomes also well determined. Fig.~\ref{fig:px} shows ${\cal P}(\beta)$ when no cuts are imposed on the kinematical variables.

\noindent
In general, the form of ${\cal P}(\beta)$ depends on the possible cuts in the kinematical variables. Still, due to the symmetric form of the possible final states, Eq.~(\ref{psiZZ}), the density matrix has a very defined structure, namely 
\bea
\rho=
\frac{1}{2+w^2}
\left(
\begin{array}{ccccccccc}
 0 & 0 & 0 & 0 & 0 & 0 & 0 & 0 & 0 \\
 0 & 0 & 0 & 0 & 0 & 0 & 0 & 0 & 0 \\
 0 & 0 & 1 & 0 & -y & 0 & 1 & 0 & 0 \\
 0 & 0 & 0 & 0 & 0 & 0 & 0 & 0 & 0 \\
 0 & 0 & -y & 0 & w^2 & 0 & -y & 0 & 0 \\
 0 & 0 & 0 & 0 & 0 & 0 & 0 & 0 & 0 \\
 0 & 0 & 1 & 0 & -y & 0 & 1 & 0 & 0 \\
 0 & 0 & 0 & 0 & 0 & 0 & 0 & 0 & 0 \\
 0 & 0 & 0 & 0 & 0 & 0 & 0 & 0 & 0 \\
\end{array}
\right)
\label{rhoxy}
\eea
with $y$ real. When one only considers final states with the same $m_1$, and $m_2$, and thus the same $\beta$, the spin state is pure and the density matrix, $\rho_{{}_\beta}$ has the form (\ref{rhoxy}) with 
$w=y=\beta$.
Otherwise $w$ and $y$ are averages over the kinematical variables, as expressed in Eq.~(\ref{average}). 
Of course, when the $\rho$ matrix is extracted from experimental data, it does not present the exact form (\ref{rhoxy}) due to systematic and statistical errors, and the existence of (small) background. 

\subsection{The irreducible tensor operator parametrization}
\label{sec:3.2}

A convenient way to parametrize the $9\times 9$ spin density-operator of the two vector bosons is to use the basis of irreducible tensor operators $\{T^{L_1}_{M_1}\otimes T^{L_2}_{M_2}\}$ \cite{Aguilar-Saavedra:2017zkn}, where
\bea
T^{L_1}_{M_1},T^{L_2}_{M_2}\in\left\{\mathbb{1}_3; T^1_1, T^1_0, T^1_{-1}; T^2_2, T^2_1, T^2_0, T^2_{-1}, T^2_{-2}\right\} \, .
\eea
Here, $T^L_M$ are normalized such that $\Tr\left\{T^L_M\; \left(T^L_M\right)^{\dagger}\right\} = 3 $, where $\left(T^L_M\right)^{\dagger}=(-1)^M \, T^L_M $. More precisely, defining $J_x, J_y$ and $J_z$ as the spin-1 components operators, we have $T^1_{\pm 1}=\mp\sqrt{3}/2\;  (J_x\pm i J_y)$ and $T^1_0=\sqrt{\frac{3}{2}}\ J_z$, i.e. 
\bea
T^1_1=\sqrt{\frac{3}{2}}
\left(
\begin{array}{ccc}
 0 & -1 & 0 \\
 0 & 0 & -1  \\
 0 & 0 & 0 \\
\end{array}
\right),
\ \ \ 
T^1_0=\sqrt{\frac{3}{2}}
\left(
\begin{array}{ccc}
 1 & 0 & 0 \\
 0 & 0 & 0  \\
 0 & 0 & -1 \\
\end{array}
\right),
\ \ \ 
T^1_{-1}=\sqrt{\frac{3}{2}}
\left(
\begin{array}{ccc}
 0 & 0 & 0 \\
 1 & 0 & 0  \\
 0 & 1 & 0 \\
\end{array}
\right).
\label{T1}
\eea
On the other hand,
\bea
T^2_{\pm 2}&=&\frac{2}{\sqrt{3}}\, (T_{\pm 1}^1)^2,
\nonumber\\
T^2_{\pm 1}&=&\sqrt{\frac{2}{3}} \left[T_{\pm 1}^1T_{0}^1 + T_{0}^1T_{\pm 1}^1\right],
\nonumber\\
T^2_0&=&\frac{\sqrt{2}}{3} \left[T_1^1T_{-1}^1 + T_{-1}^1T_{1}^1+2(T_{0}^1)^2\right].
\eea 
Explicitly,
\bea
T^2_2&=&\sqrt{3}
\left(
\begin{array}{ccc}
 0 & 0 & 1 \\
 0 & 0 & 0  \\
 0 & 0 & 0 \\
\end{array}
\right),
\ \ \ 
T^2_{-2}=\sqrt{3}
\left(
\begin{array}{ccc}
 0 & 0 & 0 \\
 0 & 0 & 0  \\
 1 & 0 & 0 \\
\end{array}
\right),
\ \ \ 
T^2_{1}=\sqrt{\frac{3}{2}}
\left(
\begin{array}{ccc}
 0 & -1 & 0 \\
 0 & 0 & 1  \\
 0 & 0 & 0 \\
 \end{array}
\right),
\nonumber\\
T^2_{-1}&=&\sqrt{\frac{3}{2}}
\left(
\begin{array}{ccc}
 0 & 0 & 0 \\
 1 & 0 & 0  \\
 0 & -1 & 0 \\
 \end{array}
\right),
\ \ \ 
T^2_0=\frac{1}{\sqrt{2}}
\left(
\begin{array}{ccc}
 1 & 0 & 0 \\
 0 & -2 & 0  \\
 0 & 0 & 1 \\
\end{array}
\right).
\label{T2}
\eea
Hence, the spin density matrix of the two vector bosons can be parametrized as
\bea
\rho=\frac{1}{9}\left[
\mathbb{1}_3\otimes \mathbb{1}_3+A^1_{LM}\ T^L_{M}\otimes \mathbb{1}_3 + A^2_{LM}\ \mathbb{1}_3\otimes T^L_{M}
+C_{L_1 M_1 L_2 M_2}\ T^{L_1}_{M_1}\otimes T^{L_2}_{M_2}
\right],
\label{rhoAC}
\eea
where we are summing in $L= 1, 2$ and $-L\leq M\leq L$ (likewise with $L_{1,2}$ and $M_{1,2}$). In order for $\rho$ to be hermitian, and taking into account that $\left(T^L_M\right)^{\dagger}=(-1)^M \, T^L_M $, the coefficients of the expansion must fulfill $A^{1,2}_{LM}=(-1)^M(A^{1,2}_{L,-M})^*$ and $C_{L_1 M_1 L_2 M_2}=(-1)^{M_1+M_2}(C_{L_1, -M_1, L_2, -M_2})^*$.  Altogether these are the 80 independent real parameters of the $9\times 9$ $\rho$ matrix.

The great advance of this parametrization over e.g. the one given by the Gell-Mann matrices, is that it allows to easily extract the values of the $A$ and $C$ coefficients from angular distributions of the final leptons. The decay density matrix of a $Z$ boson into charged leptons is given by \cite{Rahaman:2021fcz}
\bea
\Gamma=\frac{1}{4}
\left(
\begin{array}{ccc}
 1+\cos^2\theta-2\eta_\ell \cos\theta & \frac{1}{\sqrt{2}} (\sin 2\theta -2\eta_\ell\sin\theta)e^{i\varphi}& (1-\cos^2\theta) e^{i2\varphi}\\
 \frac{1}{\sqrt{2}}(\sin2\theta-2\eta_\ell \sin\theta) e^{-i\varphi}& 2\sin^2\theta& -\frac{1}{\sqrt{2}} (\sin 2\theta +2\eta_\ell\sin\theta)e^{i\varphi}  \\
 (1-\cos^2\theta) e^{-i2\varphi} & -\frac{1}{\sqrt{2}} (\sin 2\theta +2\eta_\ell\sin\theta)e^{-i\varphi} & 1+\cos^2\theta-2\eta_\ell \cos\theta \\
\end{array}
\right),\ 
\eea
where $\theta, \phi$ are the polar angles of the momentum of the negative charged lepton in the reference system in which the $Z$ is at rest, and
\begin{equation}
\eta_\ell = \frac{1-4 s_W^2}{1-4 s_W^2 + 8 s_W^4} \simeq 0.13 \,,
\label{ec:etal}
\end{equation}
with $s_W$ the sine of the electroweak mixing angle~\cite{Aguilar-Saavedra:2017zkn}. The differential cross section $ZZ\rightarrow \ell_1^+\ell_1^- \ell_2^+\ell_2^-$ is given by
\begin{equation}
\frac{1}{\sigma}\frac{d\sigma}{d\Omega_1 d\Omega_2} =
\left(\frac{3}{4\pi}\right)^2 \sum_{i, j, a, b = 1}^3 \rho_{i a,j b} (\Gamma_1)_{i j} (\Gamma_2)_{a b}
= \left(\frac{3}{4\pi}\right)^2 \Tr\left\{ \rho\; (\Gamma_1\otimes\Gamma_2)^T\right\}\ ,
\label{sigma}
\end{equation}
where $\Gamma_j=\Gamma(\theta \to \theta_j,\varphi \to \varphi_j)$ for $j=1,2$.  Using
\bea
\Tr\left\{ \mathbb{1}_3\; \Gamma^T\right\} = 2\sqrt{\pi} \ Y_0^0(\theta, \varphi),\ \ 
\Tr\left\{ T^1_M\; \Gamma^T\right\}= -\sqrt{2\pi}\eta_\ell \ Y_1^M(\theta, \varphi),
\ \ 
\Tr\left\{ T^2_M\; \Gamma^T\right\} = \sqrt{\frac{2\pi}{5}}\  Y_2^M(\theta, \varphi)
\eea
the fully differential distribution, c.f. (\ref{sigma}) can be writen in a very compact form (the sum in the $L$ and $M$ indices is implicit):
\bea
\frac{1}{\sigma}\frac{d\sigma}{d\Omega_1d\Omega_2}
&=&\frac{1}{(4\pi)^2}\left[ 1 +A_{LM}^1 B_L Y_L^M(\theta_1, \varphi_1) + A_{LM}^2 B_L Y_L^M(\theta_2, \varphi_2)\phantom{\frac{}{}}\right.
\\
&&
\left.
+ C_{L_1M_1L_2M_2} B_{L_1}B_{L_2} Y_{L_1}^{M_1}(\theta_1, \varphi_1)Y_{L_2}^{M_2}(\theta_2, \varphi_2)
 \right] \,,
\label{ec:dist4D}
\eea
with 
\bea
B_1=-\sqrt{2\pi} \eta_\ell\ , \ \ \ B_2=\sqrt{\frac{2\pi}{5}} \, .
\eea
Now, using the appropriate spherical harmonics as integration kernels, one can derive the values of the various coefficients in Eq.~(\ref{rhoAC}) from the differential cross section, namely
\bea
\int \frac{1}{\sigma}\frac{d\sigma}{d\Omega_1d\Omega_2} Y_L^M(\Omega_j)d\Omega_j &=& \frac{B_L}{4\pi} A^j_{LM},\ \ \ \ \ \ \  j=1,2 \, .
\nonumber\\
\int \frac{1}{\sigma}\frac{d\sigma}{d\Omega_1d\Omega_2} Y_{L_1}^{M_1}(\Omega_1) Y_{L_2}^{M_2}(\Omega_2)d\Omega_1 d\Omega_2&=& \frac{B_{L_1} B_{L_2}}{(4\pi)^2} C_{L_1M_1L_2M_2} \, ,
\label{ec:intY}
\eea
where we have applied the orthonormality properties of spherical harmonics.

Notice that the theoretical form of the density matrix (\ref{rhoxy}) imposes strong constraints on the various $A^j_{LM}, C_{L_1M_1L_2M_2}$ coefficients. At the end of the day it simply reads
\bea
\rho=
\left(
\begin{array}{ccccccccc}
 0 & 0 & 0 & 0 & 0 & 0 & 0 & 0 & 0 \\
 0 & 0 & 0 & 0 & 0 & 0 & 0 & 0 & 0 \\
 0 & 0 & \frac{1}{6} \left(\sqrt{2} A^1_{2,0}+2\right) & 0 & \frac{1}{3} C_{2,1,2,-1} & 0 & \frac{1}{3} C_{2,2,2,-2} & 0 & 0 \\
 0 & 0 & 0 & 0 & 0 & 0 & 0 & 0 & 0 \\
 0 & 0 & \frac{1}{3} C_{2,1,2,-1} & 0 & \frac{1}{3} \left(1-\sqrt{2} A^1_{2,0}\right) & 0 & \frac{1}{3} C_{2,1,2,-1} & 0 & 0 \\
 0 & 0 & 0 & 0 & 0 & 0 & 0 & 0 & 0 \\
 0 & 0 & \frac{1}{3} C_{2,2,2,-2} & 0 & \frac{1}{3} C_{2,1,2,-1} & 0 & \frac{1}{6} \left(\sqrt{2} A^1_{2,0}+2\right) & 0 & 0 \\
 0 & 0 & 0 & 0 & 0 & 0 & 0 & 0 & 0 \\
 0 & 0 & 0 & 0 & 0 & 0 & 0 & 0 & 0 \\
\end{array}
\right) \, .
\label{rhomatrixAC}
\eea
with 
\be
\frac{1}{\sqrt{2}} A^1_{2,0}+1 =C_{2,2,2,-2}.
\label{ligadura}
\ee
We do not replace the latter relation in (\ref{rhomatrixAC}). It could be used, for example, as a way to estimate the uncertainties in the experimental determination of the density matrix, or to improve the determination of the independent coefficients and thereby improve the precision in the measurement of the entanglement observables. An investigation of the optimal way to extract the latter from data is beyond the scope of the present work.

\section{Conditions for entanglement} \label{sec:EntConds}

Intuitively, a ``classical" system of two vector bosons with vanishing spin-third-component can only be in three states: $\ket{+ -}, \ket{0 0}$ or $\ket{- +}$. Any superposition of these possibilities implies an entangled quantum state. Hence, one can expect that if the $\rho-$matrix is {\em non}-entangled, it can contain just three non-vanishing entries, namely the diagonal ones:  $\rho_{+-,+-}$, $\rho_{00,00}$ and $\rho_{-+,-+}$. Thus, if any of the six remaining entries is different from zero, that would be a signal of entanglement. It is interesting to show that this is indeed the case, by using the above-mentioned Peres-Horodecki criterion, see Eq.~(\ref{rhoT2}) and below. For a generic spin-density matrix with vanishing third-component,
\bea
\rho=
\left(
\begin{array}{ccccccccc}
 0 & 0 & 0 & 0 & 0 & 0 & 0 & 0 & 0 \\
 0 & 0 & 0 & 0 & 0 & 0 & 0 & 0 & 0 \\
 0 & 0 & a & 0 & b & 0 & c & 0 & 0 \\
 0 & 0 & 0 & 0 & 0 & 0 & 0 & 0 & 0 \\
 0 & 0 & b^* & 0 & d & 0 & f & 0 & 0 \\
 0 & 0 & 0 & 0 & 0 & 0 & 0 & 0 & 0 \\
 0 & 0 & c^* & 0 & f^* & 0 & g & 0 & 0 \\
 0 & 0 & 0 & 0 & 0 & 0 & 0 & 0 & 0 \\
 0 & 0 & 0 & 0 & 0 & 0 & 0 & 0 & 0 \\
\end{array}
\right) \, ,
\label{rhogeneric}
\eea
the corresponding partially transposed matrix reads
\bea
\rho=
\left(
\begin{array}{ccccccccc}
 0 & 0 & 0 & 0 & 0 & 0 & 0 & 0 & c \\
 0 & 0 & 0 & 0 & 0 & b & 0 & 0 & 0 \\
 0 & 0 & a & 0 & 0 & 0 & 0 & 0 & 0 \\
 0 & 0 & 0 & 0 & 0 & 0 & 0 & f & 0 \\
 0 & 0 & 0 & 0 & d & 0 & 0 & 0 & 0 \\
 0 & b^* & 0 & 0 & 0 & 0 & 0 & 0 & 0 \\
 0 & 0 & 0 & 0 & 0 & 0 & g & 0 & 0 \\
 0 & 0 & 0 & f^* & 0 & 0 & 0 & 0 & 0 \\
 c^* & 0 & 0 & 0 & 0 & 0 & 0 & 0 & 0 \\
\end{array}
\right) \, ,
\label{rhogenericT2}
\eea
which has eigenvalues $a,d,g, \pm |b|, \pm |c|, \pm|f|$. Therefore if $b\neq 0$, $c\neq 0$ or $f\neq 0$ the density matrix is entangled. Note that the reverse is also true: if $b=c=f=0$ the state is obviously separable, as $\rho$ is diagonal in the separable basis. This represents a noteworthy example beyond a two-qubit system, where, thanks to an underlying symmetry, the Peres-Horodecki condition for entanglement is not just sufficient, but also necessary.

When applied this condition to our density matrix (\ref{rhomatrixAC}), it turns out that the $ZZ$ system is entangled  if and only if
\bea
C_{2,1,2,-1}\neq 0\ \ \ \ \ \ \ {\rm or}\ \ \ \ \ \ \ C_{2,2,2,-2}\neq 0 \,.
\label{EntCond}
\eea

\section{Conditions for violation of Bell inequalities}\label{sec:BellConds}

As mentioned in section \ref{sec:generalditions}, the explicit form of the CGLMP inequality depends on the specific choice of the four observables, $A_1, A_2, B_1, B_2$ associated to the Alice and Bob Hilbert spaces. The optimal choice, i.e. the one that leads to a larger violation of the inequality, depends on the state at hand (in our case the density operator, $\rho$).

This issue was considered in Ref.\cite{CGLMP2002} in a more abstract context. Namely, denoting by $\ket{i}_A$, $\ket{j}_B$ ($i,j=1,2,3$) two orthonormal bases of  ${\cal H}_A$, ${\cal H}_B$, if the state at hand is the maximally entangled state of the form
\begin{equation}
\ket{\psi'} = \frac{1}{\sqrt{3}}\left(\ket{11} +  \ket{22} + \ket{33}\right) \, ,
\label{psi}
\end{equation}
where $\ket{ij}=\ket{i}_A\ket{j}_B$, then a particular choice of the four 
observables $A_1, A_2, B_1, B_2$ was argued to maximize the violation of the CGLMP inequality. A compact way to express this optimal choice is by building the corresponding Bell operator, say ${\cal O}_{\text Bell}'$ \cite{Acin:2002zz}. In terms of the $T^L_M$ matrices of Eqs.(\ref{T1}, \ref{T2}) ${\cal O}_{\text Bell}'$ reads
\begin{equation}
{\cal O}_{\text Bell}'=  
\displaystyle{\frac{4}{ 3 \sqrt{3}} }
\left( T^1_{1} \otimes T^1_{1} + T^1_{-1} \otimes T^1_{-1} \right )+
\displaystyle{\frac{2}{ 3}} \, \left( T^2_{2} \otimes T^2_{2} + T^2_{-2} \otimes T^2_{-2} \right) \, .
\label{Oprime}
\end{equation}

Coming back to the $H\rightarrow ZZ$ decay, and working in the usual 
spin basis 
\bea
\{\ket{+},\ket{0}, \ket{-}\}_A\otimes \{\ket{+},\ket{0}, \ket{-}\}_B\ ,
\label{spinbasis}
\eea
for a particular event, the spin state of the $ZZ$ system is given by Eq.~(\ref{psiZZ}), which in general does not have the form (\ref{psi}). However, in the non-relativistic limit, which corresponds to $\beta=1$ in Eq.~(\ref{psiZZ}), the  state is a pure singlet,
\bea
 \ket{\psi_s}\frac{1}{\sqrt{3}}\left(\ket{++} -  \ket{00} + \ket{--}\right)\ ,
\label{psi2}
\eea
which can be written in the form (\ref{psi}) by a change of basis defined by the unitary transformation
\bea
 \ket{\psi_s}\rightarrow  U O_A \otimes U^* \ket{\psi_s} \, ,
\eea
where 
\begin{equation}
O_A
=
\left( {\begin{array}{ccc}
 0 & 0  & 1 \\
 0 & -1 & 0\\
   1 &0 &0\\
  \end{array} } \right) 
\end{equation}
and $U$ is an arbitrary $3\times 3$ unitary matrix, $U\in U(3)$.

Hence, in the non-relativistic limit and working in this new basis, an optimal choice for the Bell operator is the operator  ${\cal O}_{\text Bell}'$, given in  Eq.~(\ref{Oprime}). Then the violation of the CGLMP inequality (\ref{I3}) reads
\bea
I_3=\Tr\left\{( U O_A \otimes U^* ) \rho
( U O_A \otimes U^* )^\dagger  {\cal O}_{\text Bell}' \right\} > 2
\label{OBell1}
\eea
where $\rho=\ket{\psi_s}\bra{\psi_s}$ is the density operator in the basis (\ref{spinbasis}). Note that $\rho$ has the form (\ref{rhoxy}) with $y=w=\beta=1$.
Then, using the invariance of the trace under cyclic permutations, the violation of the CGLMP inequality can be expressed as
\bea
I_3=\Tr\left\{ \rho\ {\cal O}_{\text Bell}\right\}> 2
\label{OBell1p}
\eea
with
\bea
{\cal O}_{\text Bell}=( U O_A \otimes U^* )^\dagger  {\cal O}_{\text Bell}'  (U O_A \otimes U^* ) \, .
\label{OBell2}
\eea
In other words, this operator represents the optimal choice for the Bell operator when the spin state is the singlet ($y=w=1$) and one is working in the physical spin basis (\ref{spinbasis}). Note that in this limit all the choices of $U$ give equivalent results for the violation of the CGLMP inequality,  Eq.~(\ref{OBell1p}).

Of course, in general $\beta\neq 1$, so the spin state $\ket{\psi_{ZZ}}$ (\ref{psiZZ}) is not a singlet. Thus
one should in principle explore all the possible Bell operators in order to find the optimal one. This represents a huge parameter space. However, since $\beta$ is typically not far from 1, see Fig.~\ref{fig:px}, a reasonable strategy is to consider Bell operators of the form (\ref{OBell2}), exploring the space of $U$-matrices (notice
that now not all choices of $U$ are equivalent). 

\begin{figure}[h!]
   \begin{center}
        \includegraphics[scale=0.45]{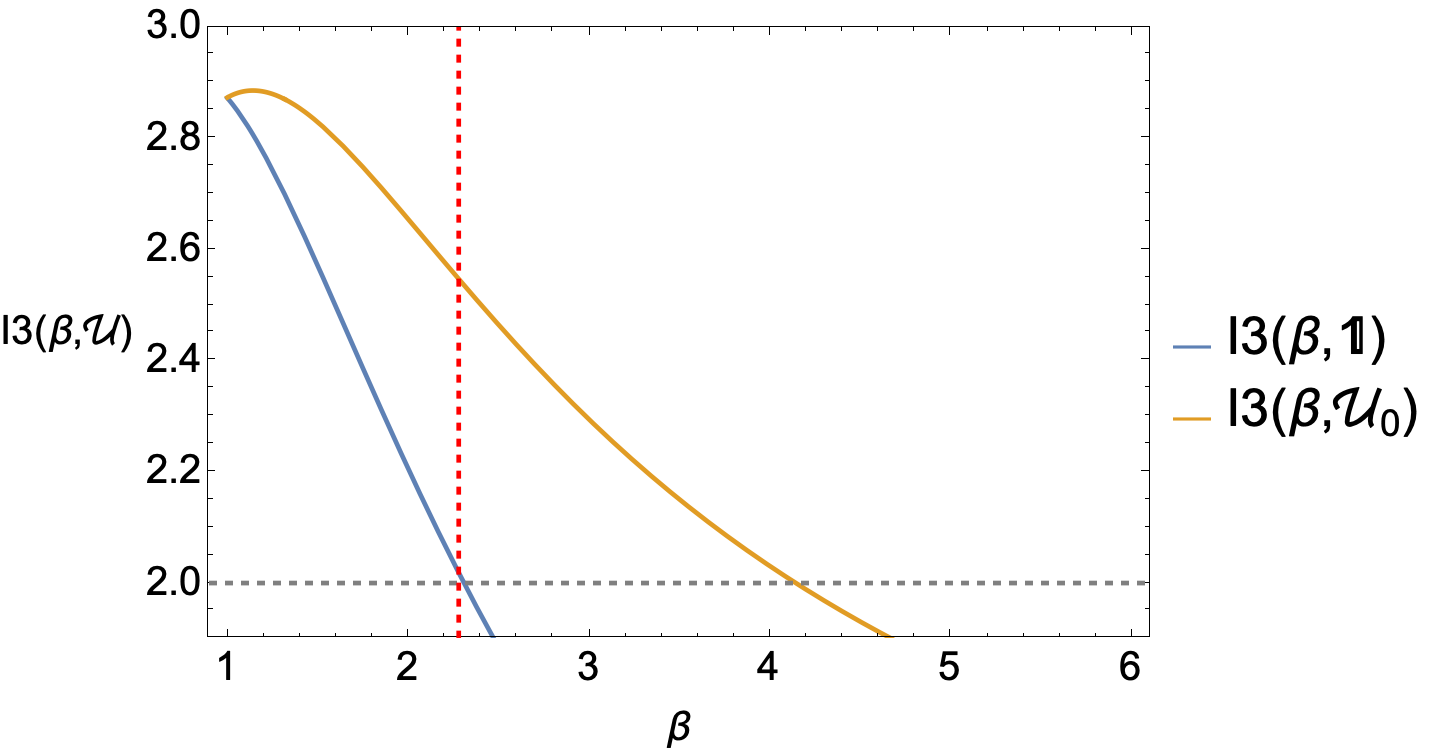}
     \end{center}
	\caption{Functions 
  $\left (I_3(\beta,  \mathbb{1} ), I_3(\beta, U_0) \right)$ defined in Eq.~(\ref{I3U}). We have also displayed the local-realistic upper bound for the Bell inequality (gray dashed horizontal line) as well as the mean value of $\beta$ with respect to the probability distribution ${\cal P}(\beta)$ shown in Fig. \ref{fig:px}  (red dashed vertical line).}
    \label{fig:px2}
\end{figure}

This exploration is greatly simplified by noting that 
the $\ket{\psi_{ZZ}}$ state (\ref{psiZZ}) is still invariant under transformations $UO_A\otimes U^*$, where $U$ is a unitary matrix acting only in the subspace spanned by $\{\ket{+}, \ket{-}\}$, as well as under a redefinition via a global phase. Hence, the non-equivalent choices of $U$ are those belonging to the $U(3) / \left(U(2)\otimes U(1)\right)$ coset. 
For each one of them we get a different ${\cal O}_{\text Bell}$ operator (\ref{OBell2}) and a prediction for $I_3=\langle{\cal O}_{\text Bell}\rangle$, say $I_3(\beta, U)$. We have explored the form of $U$ that maximizes $I_3(\beta, U)$ and found that, in the region of interest, a convenient choice is:
\begin{equation}
U_0
 =
\left(
  {\begin{array}{ccc}
  \frac{1}{2} & -  \frac{1}{\sqrt 2 }  &   \frac{1}{2} \\
 -  \frac{1}{\sqrt 2 }  & 0 & \frac{1}{\sqrt 2 } \\
   \frac{1}{2} & \frac{1}{\sqrt 2 } &  \frac{1}{2}\\
  \end{array} }
   \right)\ .
\end{equation}
This is illustrated in Fig.~\ref{fig:px2}, where the functions
\begin{equation}
\begin{array}{ccl}
I_3(\beta,  \mathbb{1} )  & =  & \displaystyle { 
\frac{12 + 8 \sqrt{3}\beta }{3 (2+ \beta^2 )} },   \\[4mm]
I_3(\beta, U_0)  & =  & \displaystyle { 
\frac{(1+\beta)(3 + 4 \sqrt{3}+ 3\beta) }{3 (2+ \beta^2 )}} , 
\end{array}
\label{I3U}
\end{equation}
are displayed, showing the improvement provided by
this non-trivial choice.

For completeness, we give the expression of the Bell operator (\ref{OBell2}) for $U=U_0$:
\begin{equation}
\begin{array}{rcl}
{\cal O}_{\text Bell} & = &   
\left(
\displaystyle{\frac{2}{ 3 \sqrt{3}} } \, 
\left(
T^1_{1} \otimes T^1_{1}  -  T^1_{0} \otimes T^1_{0}  +T^1_{1} \otimes T^1_{-1}
\right) + 
\displaystyle{\frac{1}{ 12} }\left(T^2_{2} \otimes T^2_{2} +T^2_{2} \otimes T^2_{-2}\right) \right.
 \\[5mm] 
 && \left.
+\displaystyle{\frac{1}{ 2\sqrt{6}}} \, \left( T^2_{2} \otimes T^2_{0} +  T^2_{0} \otimes T^2_{2}\right) - 
\displaystyle{\frac{1}{ 3} }\, ( T^2_{1} \otimes T^2_{1} +  T^2_{1} \otimes T^2_{-1} )+ \displaystyle{\frac{1}{ 4} }\, T^2_{0} \otimes T^2_{0}  
\right) 
 + h.c.
\end{array}
\end{equation}
In general, the state of the $ZZ$ system is given by the density operator $\rho$ shown in Eq.~(\ref{rhomatrixAC}) and the corresponding prediction for 
$I_3=\Tr \left\{\rho \;  {\cal O}_{\text Bell} \right\}$ reads
\begin{equation}
I_3 = \frac{1}{36} 
\left( 
18 + 16 \sqrt{3} - \sqrt{2} \left(9- 8   \sqrt{3}  \right) A^1_{2,0} - 8 \left(3 +  2   \sqrt{3}  \right) C_{2,1,2,-1}  + 6\, C_{2,2,2,-2}
\right) \, .
\label{BellCond}
\end{equation}
In other words, 
the $ZZ$ system violates the GCLMP inequality whenever this expression for $I_3$ is larger than 2.
Notice that the $A^1_{2,0}$ parameter is related to $C_{2,2,2,-2}$ by Eq.~(\ref{ligadura}). However, we will keep it as an independent parameter when extracting its value from  data, as a possible handle to estimate the involved uncertainties\footnote{Actually, in this way the statistical uncertainty in $I_3$ becomes 5\%  smaller in the simulations discussed in section \ref{sec:numerical}.}.
This may not be convenient, however, in the presence of systematic uncertainties, but such a study is beyond the scope of the present work.

\section{Numerical results}\label{sec:numerical}

We investigate now the statistical sensitivity of future experimental measurements to the above described entanglement and violation of Bell inequalities. For this goal we generate $pp \to H \to ZZ^* \to 4\ell$ using {\scshape MadGraph}~\cite{Alwall:2014hca}, with on-shell $H$. The $gg \to H$ one-loop process is implemented by a contact interaction, but otherwise the Monte Carlo calculation is performed at the leading order. On the other hand, for the calculation of  the expected number of events we use state-of-the art values of the Higgs production cross section and branching ratio into four electrons or muons. The cross section at next-to-next-to-next-to-leading order is 48.61 pb at a centre-of-mass energy of 13 TeV~\cite{Cepeda:2019klc}, and the Higgs branching ratio decay into four leptons (electrons or muons) is $1.24 \times 10^{-4}$~\cite{Cepeda:2019klc}. Overall, the cross section times branching ratio for the final state considered is 6.02 fb at 13 TeV. We implement our analysis using the $e^+ e^- \mu^+ \mu^-$ final state for simplicity, but note that in final states with four electrons or four muons the two $Z$ bosons can still be cleanly identified, since one is nearly on-shell and gives two leptons with invariant mass close to $m_Z$, while the remaining two leptons have a much lower invariant mass. The (very close to real) $Z$ boson with largest invariant mass is labelled as $Z_1$, and its four-momentum reconstructed from its decay products $\ell_1^+ \ell_1^-$. The off-shell $Z$ boson is labelled as $Z_2$, and its momentum is determined summing up the momenta of its decay products $\ell_2^+ \ell_2^-$.

The reference system used to determine spin observables has a construction similar to the helicity basis used for top pair production~\cite{Bernreuther:2015yna}, and is defined as follows:
\begin{itemize}
\item The $\hat z$ axis is taken in the direction of the $Z_1$ three-momentum in the Higgs boson rest frame.
\item The  $\hat x$ axis is in the production plane and defined as $\hat x = \mathrm{sign}(\cos \theta) (\hat p_p - \cos \theta \hat z)/\sin \theta$, with $\hat p_p = (0,0,1)$ the direction of one proton in the laboratory frame, $\cos \theta = \hat z \cdot \hat p_p$. The definition for $\hat x$  is the same if we use the direction of the other proton $- \hat p_p$. 
\item The $\hat y$ axis is taken such that $\hat y = \hat z \times \hat x$, orthogonal to the production plane.
\end{itemize}
The angles $(\theta_1,\varphi_1)$ in the differential distribution (\ref{ec:dist4D}) are the polar coordinates of the three-momentum of the negative charged lepton $\ell_1^-$ from the $Z_1$ decay, in the $Z_1$ rest frame. Likewise, the angles $(\theta_2,\varphi_2)$ correspond to the three-momentum of $\ell_2^-$ in the $Z_2$ rest frame. As mentioned in section~\ref{sec:3.2}, the coefficients of the expansion of the $ZZ$ density operator (\ref{rhoAC}) are obtained by integration with spherical harmonics, c.f. (\ref{ec:intY}).

We estimate the potential of future measurements by calculating the statistical uncertainty of the relevant observables for entanglement and Bell inequality violation,  see Eqs.~(\ref{EntCond}), (\ref{BellCond}). For the LHC Runs $2+3$ and the HL-LHC the luminosities assumed are of 300 fb$^{-1}$ and 3 ab$^{-1}$, respectively. For simplicity, a CM energy of 13 TeV is used in all cases; note however that the cross section for $gg \to H$ increases to 54.7 pb at 14 TeV. In order to have a more realistic estimate of the number of events in each case a lepton detection efficiency of 0.7 is assumed, yielding an overall detection efficiency of 0.25.\footnote{ This efficiency accounts for the minimum transverse momentum ($p_T$) thresholds required for lepton detection. We do not include any trigger requirement. The presence of four leptons from the Higgs decay, some of them with significant $p_T$, is expected to fulfill one or many of the trigger conditions for one, two, or three leptons, see e.g. Ref.~\cite{ATL-DAQ-PUB-2018-002}.}

The statistical uncertainty in the observables is determined by performing $1000$ pseudo-experiments. In each one, we select a random set of $N$ events, with a size corresponding to the assumed luminosity, and calculate the observables from the differential distribution, as aforementioned. Repeating this procedure, we obtain the mean and standard deviation for each observable. The mean value resulting from the pseudo-experiments is quite close to the theoretical value calculated with the full Monte Carlo sample, and the standard deviation corresponds to the expected statistical uncertainty. Systematic uncertainties are not included in our analysis. Given the clean final state and the good experimental resolution for charged leptons, these uncertainties are expected to be small. In any case, they must be addressed within an experimental analysis using a full detector simulation.

As discussed in section~\ref{sec:3.1}, the larger the mass $m_2$ of the off-shell $Z$ boson, the more entangled the $ZZ$ state is. However, requiring a lower cut on $m_{Z_2}$ also decreases the statistics, increasing the uncertainty in the measurements. We give results without any cut and also with lower cuts $m_{Z_2} \geq 10,20,30$ GeV. 

Table~\ref{tab:LHC} gives the results for $L = 300~\text{fb}^{-1}$. The entanglement can be probed at the $3\sigma$ level using $C_{2,1,2,-1}$, and below the $2\sigma$ level using $C_{2,2,2,-2}$, see Eq.~(\ref {EntCond}). A combination of both observables, which is beyond the scope of this work, would improve the sensitivity. On the other hand, the sensitivity to the violation of the Bell inequalities, see Eq.~(\ref{BellCond}), is below the $2\sigma$ level.

\begin{table}[htb]
\begin{center}
\begin{tabular}{cccccc}
               & \multicolumn{4}{c}{min $m_{Z_2}$} \\
               & 0                & 10 GeV           & 20 GeV           & 30 GeV \\ \hline
$N$            & 450              & 418              & 312              & 129    \\
$C_{2,1,2,-1}$ & $-0.98 \pm 0.31$ & $-0.97 \pm 0.33$ & $-1.05 \pm 0.38$ & $-1.06 \pm 0.61$
\\
$C_{2,2,2,-2}$ & $0.60 \pm 0.37$  & $0.64 \pm 0.38$  & $0.74 \pm 0.43$ & $0.82 \pm 0.63$
\\
$I_3$          & $2.66 \pm 0.46$  & $2.67 \pm 0.49$  & $2.82 \pm 0.57$ & $2.88 \pm 0.89$
\end{tabular}
\caption{Values of the spin correlation coefficients $C_{2,1,2,-1}$ and $C_{2,2,2,-2}$ signaling quantum entanglement, and the Bell operator $I_3$ signaling violation of the Bell inequalities, obtained from 1000 pseudo-experiments with with $L = 300~\text{fb}^{-1}$.}
\label{tab:LHC}
\end{center}
\end{table}

Table~\ref{tab:HLLHC} gives the results for $L = 3~\text{ab}^{-1}$. In this case, the entanglement can be probed beyond the $5\sigma$ level using both coefficients, reaching a $10\%$ precision in the case of $C_{2,1,2,-1}$. The sensitivity to a violation of the Bell inequalities is at the $4.5\sigma$ level.

\begin{table}[htb]
\begin{center}
\begin{tabular}{cccccc}
               & \multicolumn{4}{c}{min $m_{Z_2}$} \\
               & 0                & 10 GeV           & 20 GeV           & 30 GeV \\ \hline
$N$            & 4500             & 4180             & 3120             & 1290    \\
$C_{2,1,2,-1}$ & $-0.95 \pm 0.10$ & $-1.00 \pm 0.10$ & $-1.04 \pm 0.12$ & $-1.04 \pm 0.19$
\\
$C_{2,2,2,-2}$ & $0.60 \pm 0.12$  & $0.64 \pm 0.12$  & $0.74 \pm 0.14$ & $0.83 \pm 0.20$
\\
$I_3$          & $2.63 \pm 0.15$  & $2.71 \pm 0.16$  & $2.81 \pm 0.18$ & $2.84 \pm 0.28$
\end{tabular}
\caption{The same as Table~\ref{tab:LHC}, for $L = 3~\text{ab}^{-1}$.}
\label{tab:HLLHC}
\end{center}
\end{table}

\section{Summary and conclusions}
\label{sec:conclusions}

In this paper we have studied the quantum properties of the $ZZ$ pair produced in the Higgs boson decay, focusing on entanglement and violation of Bell inequalities. Because the Higgs boson is a scalar and $m_Z$ is not far from  $m_H$, the $ZZ$ pair is produced nearly in a spin-singlet state, where the quantum entanglement is maximal, and turns the study of its quantum properties quite interesting. 

The spin state of a $ZZ$ pair, and the joint decay angular distribution, are parameterised by 80 (polarization and spin correlation)
observables. 
We have introduced a novel, compact parameterisation that makes the description strikingly simple. For the specific case of $H \to ZZ$, angular momentum, and P and CP conservation, greatly reduce the number of parameters to only two  independent parameters, namely the spin correlation coefficients $C_{2,1,2,-1}$ and $C_{2,2,2,-2}$, see Eq.~(\ref{rhomatrixAC}).

In terms of these parameters we have first formulated the conditions for quantum entanglement of the $ZZ$ pair. There are two sufficient  Peres-Horodecki conditions for entanglement, namely that any of the coefficients, $C_{2,1,2,-1}$ or $C_{2,2,2,-2}$, is different from zero.
We have shown that, due to the underlying symmetry of the process, these are also necessary conditions, which represents a noteworthy exception to the general situation for the Peres-Horodecki criterion.
Second, we
have found a Bell operator which is a much more powerful indicator of the violation of Bell inequalities than other proposals in the literature
when $ZZ$ is not exactly in a spin-singlet state (as it is the actual case) .

Finally, we have investigated the experimental prospects to measure the entanglement and violation of Bell inequalities, focusing on the gluon-fusion process $gg \to H \to ZZ \to 4 \ell$ at the LHC. Our study has been performed at the parton level, without a detector simulation and without considering the background. Concerning the former, we do not expect a great impact of the detector resolution, which is quite good for charged leptons at the ATLAS and CMS detectors. By the same token, systematic uncertainties are expected to be small. The experimental precision is then expected to be dominated by the available statistics: the decay $H \to 4 \ell$ has a branching ratio of only $1.24 \times 10^{-4}$. 

Regarding the backgrounds, the $H \to ZZ \to 4 \ell$ signal is quite clean, its main background being the electroweak process $pp \to ZZ/Z\gamma \to 4 \ell$ which is about 4 times smaller at the Higgs peak~\cite{CMS:2021ugl} 
Yet, a background subtraction will be necessary before computing the entanglement observables. This non-negligible background will slightly increase the statistical uncertainty of the measurement.

Another issue to be taken into account is the fact that in four-electron and four-muon final states, LO interference diagrams from the interchange of identical particles give non-negligible contributions that are up to order 10\% in some distributions~\cite{LHCHiggsCrossSectionWorkingGroup:2013rie}. These contributions may be reduced when a kinematical cut on the on-shell $Z$ mass is imposed, for example, and do not invalidate the interpretation of the measurements as corresponding to entanglement or violation of Bell inequalities.

 The decay $H \to WW \to \ell \nu \ell \nu$ considered in  \cite{Barr:2021zcp} shares quite the same properties of $H \to ZZ$; however, a unique kinematical reconstruction of the final state is not possible due to the two unobserved neutrinos. Even if the momenta of the neutrinos might be {\em guessed} with a kinematical fit, there would be a continuum of possible solutions because there are fewer kinematical constraints than unknowns. This fact may compromise the observability of the entanglement and violation of Bell inequalities by measurements of spin and spin correlation coefficients in this decay, making $H \to ZZ$ quite unique.

With these caveats in mind, there are good prospects for the measurements at the LHC. With the Run 2+3 combined luminosities, the quantum entanglement might be measured at the $3\sigma$ level. At the HL-LHC, the significance would be well above $5 \sigma$, and the violation of Bell inequalities would be probed at the $4.5\sigma$ level.

\vspace{0.3cm}

\section*{Acknowledgements}

We are grateful to  J. Bernabeu  and C. Gonz\'alez-Garc\'{\i}a for useful discussions.
This work is supported by the grants IFT Centro de Excelencia Severo Ochoa SEV-2016-0597, CEX2020-001007-S and by PID2019-110058GB-C21  and PID2019-110058GB-C22 funded by MCIN/AEI/10.13039/501100011033 and by ERDF. The work of A.B. is supported through the FPI grant PRE2020-095867 funded by MCIN/AEI/10.13039/501100011033, and by FCT project CERN/FIS-PAR/0004/2019.

\appendix
\section{General construction of the Bell operator for the CGLMP inequality}
\label{sec:AppendixA}

The CGLMP inequality \cite{CGLMP2002} is a testable consequence of local realism, which is more powerful than the well-known CHSH inequalities \cite{Clauser:1969ny}. 
Suppose that the quantum system consists of two 
two subsystems (Alice and Bob), both with dim-$d$ Hilbert space. Let us denote $A_a$ and $B_b$, with $a,b=1,2$, the operators measured respectively by Alice and Bob. They are assumed to have non-degenerate eigenvalues $0, 1, \cdots , d-1$. Then
the CGLMP inequality reads
\bea
I_d\leq 2
\eea
with 
\bea
I_d 
&=&
\sum_{k=0}^{[d/2]-1}
\left( 1-\frac{2k}{d-1}\right)
\left(
P\left(A_1=B_1+k\right)+P\left(B_1=A_2+k+1\right)+P\left(A_2=B_2+k\right)\phantom{\frac{1}{2}}
\right.
\nonumber\\
&&
+P\left(B_2=A_1+k\right)-\left[P\left(A_1=B_1-k-1\right)
+P\left(B_1=A_2-k\right)
\phantom{\frac{1}{2}}
\right.
\nonumber
\\
&&
\left.\left.\phantom{\frac{1}{2}}
+P\left(A_2=B_2-k-1\right)+P\left(B_2=A_1-k-1\right)\right]
\right)\, ,
\label{CGLMP2}
\eea
where $P\left(B_i=A_j+a\right)$ denotes the probability that the measurement of the observable $B_i$ gives the same result as the one of $A_j$ plus $a$ (mod. $d$). 

The $A_a$ and $B_b$ observables can be defined by their respective  normalized eigenstates, $\ket{k}_{A_a}$ and $\ket{l}_{B_b}$, which can in turn be specified by their components in certain bases
(computational bases), say $\{\ket{\alpha}_A\}$ and $\{\ket{\beta}_B\}$, with $\alpha, \beta=1,\cdots , d$.
Those components can be arranged as the columns of four unitary matrices, $U_{A_a},\, U_{B_b}$. Thus, specifying the unitary matrices $\left(U_{A_a},\, U_{B_b}\right)$
is equivalent to give the Hermitian
 operators $\left(A_a,B_b \right)$.
 
 Now, in quantum mechanics the probability of measuring a certain outcome is given by the expectation value of the projector associated to that eigenvalue. Hence, the quantum prediction for $I_d$ is given by the expectation value of an certain operator,
 \be
 I_d=\langle{\cal O}_{\text Bell}\rangle\ ,
 \ee
 which is a cumbersome combination of the projectors involved in the probabilities of Eq.~(\ref{CGLMP2}). After some algebra, the Bell operator can be expressed as
\begin{equation}
\begin{array}{rcl}
{\cal O}_{\text Bell}&=& \dfrac{2}{d-1}\left\{-\left[U_{A_1}\otimes U_{B_1}\right] P_1 \left[\mathbb{1}_d \otimes J_z\right] P_1^\dagger \left[U_{A_1}\otimes U_{B_1}\right]^\dagger \right. \\[3mm] 

&& \left. +\left[U_{A_1}\otimes U_{B_2}\right] P_0 \left[\mathbb{1}_d \otimes J_z\right] P_0^\dagger \left[U_{A_1}\otimes U_{B_2}\right]^\dagger\right. \\[3mm] 

&& \left. +\left[U_{A_2}\otimes U_{B_1}\right] P_1 \left[\mathbb{1}_d \otimes J_z\right] P_1^\dagger \left[U_{A_2}\otimes U_{B_1}\right]^\dagger\right. \\[3mm] 

&& \left. -\left[U_{A_2}\otimes U_{B_2}\right] P_1 \left[\mathbb{1}_d \otimes J_z\right] P_1^\dagger \left[U_{A_2}\otimes U_{B_2}\right]^\dagger\right\},
\end{array}
\end{equation}
where $\mathbb{1}_d$ is the $d\times d$ identity matrix. The $d\times d$ matrix $J_z$ is the spin-$J$ third component operator, where we have identified $J= \dfrac{d-1}{2}$:

\begin{equation}
J_z= \left({\begin{array}{cccc}
J       & 0         & \cdots    & 0 \\
0       & J -1      & \cdots    & 0 \\
\vdots  & \vdots    & \ddots    & \vdots \\
0       & 0         & \cdots    & -J
\end{array} } \right).
\end{equation}

The $d^2\times d^2 $ matrices $P_n$, $n=0,1$ (in general $n=0,\dots,d-1$) are block-diagonal permutation matrices

\begin{equation}
P_n= \left({\begin{array}{ccccc}
C^{\,n}     & \mathscr{O}& \cdots      & \mathscr{O}  \\
\mathscr{O} & C^{\,n+1}  & \cdots      & \mathscr{O}  \\
\vdots      & \vdots     & \ddots      & \vdots \\
\mathscr{O} & \mathscr{O}& \mathscr{O} & C^{\,n+(d-1)}
\end{array} } \right),
\end{equation}
where $\mathscr{O}$ is the $d\times d$ null matrix and $C$ is the 
$d\times d$ cyclic permutation 
\begin{equation}
C= \left({\begin{array}{ccccc}
0       & 0      & \cdots  & 0      & 1  \\
1       & 0      & \cdots  & 0      & 0  \\
0       & 1      & \ddots  & \vdots & \vdots  \\
\vdots  & \vdots & \ddots  & 0      & 0  \\
0       & 0      & \cdots  & 1      & 0  
\end{array} } \right).
\end{equation}

\vspace{0.2cm}
\noindent
For the case examined in this paper, $d=3$, we have chosen $A_a, B_b $ such that their associated $U_{A_a,B_b}$ matrices are:

\begin{equation}
\begin{array}{rcll}
U_{A_1}  &= & O_A \, U_0 & U_{ME,\,A_1}\,, \\
U_{A_2}  &= & O_A \, U_0 & U_{ME,\,A_2}\,, \\
U_{B_1}  &= & \phantom{O_A \,} U_0   & U_{ME,\,B_1} \,, \\
U_{B_2}  &= & \phantom{O_A \,} U_0   & U_{ME,\,B_2} \,, 
\end{array}
\end{equation}
where $U_{ME,\,A_a}$ and $U_{ME,\,B_b}$ are the unitary matrices associated with the Bell operator that is supposed to be optimal for the maximally entangled state \cite{CGLMP2002,Acin:2002zz}. In particular:

\begin{equation}
\begin{array}{rcll}
U_{ME,\,A_1}  &= & U_D^{0} & U_{FT} \,, \\
U_{ME,\,A_2}  &= & U_D^{\frac{1}{2}} & U_{FT} \,, \\
U_{ME,\,B_1}  &= & U_D^{\frac{1}{4}} & U_{FT}^* \,, \\
U_{ME,\,B_2}  &= & U_D^{-\frac{1}{4}}& U_{FT}^* \,,
\end{array}
\end{equation}
with

\begin{equation}
U_{FT}
=
\left( {\begin{array}{ccc}
  1 & 1  & 1 \\
   1 &  \omega & \omega^2 \\
   1 &  \omega^2 & \omega^4 \\
  \end{array} } \right) \, ,  \;\;\;\;
U_D^\alpha= 
\left( {\begin{array}{ccc}
  1 & 0  & 0 \\
   0 &  \omega^\alpha & 0 \\
  0 &  0 &  \omega^{2 \alpha} \\
  \end{array} } \right) 
  \end{equation}
and $\omega = e^{i\frac{2\pi}{3}}$.
Finally,
\begin{equation}
O_A
=
\left( {\begin{array}{ccc}
 0 & 0  & 1 \\
 0 & -1 & 0\\
   1 &0 &0\\
  \end{array} } \right) \, ,  \;
  U_0
 =
\left(
  {\begin{array}{ccc}
  \frac{1}{2} & -  \frac{1}{\sqrt 2 }  &   \frac{1}{2} \\
 -  \frac{1}{\sqrt 2 }  & 0 & \frac{1}{\sqrt 2 } \\
   \frac{1}{2} & \frac{1}{\sqrt 2 } &  \frac{1}{2}\\
  \end{array} }
   \right)\ .
  \end{equation}

\bibliographystyle{style1.bst}
\bibliography{main.bib}

\end{document}